# Interface Optimization via Fullerene Blends Enables Open-Circuit Voltages of 1.35 V in $CH_3NH_3Pb(I_{0.8}Br_{0.2})_3$ Solar Cells

*Zhifa Liu[#], Johanna Siekmann[#§], Benjamin Klingebiel, Uwe Rau, and Thomas Kirchartz*[§]*

Dr. Z. Liu, J. Siekmann, Dr. B. Klingebiel, Prof. U. Rau, Prof. T. Kirchartz
IEK5-Photovoltaik, Forschungszentrum Jülich, 52425 Jülich, Germany

\# Author Contributions
Z.L. and J.S. contributed equally.

*Prof. T. Kirchartz
Faculty of Engineering and CENIDE, University of Duisburg-Essen, Carl-Benz-Str. 199, 47057 Duisburg, Germany

§ Corresponding Authors
E-mail: t.kirchartz@fz-juelich.de
E-mail: j.siekmann@fz-juelich.de

Non-radiative recombination processes are the biggest hindrance to approaching the radiative limit of the open-circuit voltage for wide-band gap perovskite solar cells. In addition, to high bulk quality, good interfaces and good energy level alignment for majority carriers at charge transport layer-absorber interfaces are crucial to minimize non-radiative recombination pathways. By tuning the lowest-unoccupied molecular-orbital of electron transport layers via the use of different fullerenes and fullerene blends, we demonstrate open-circuit voltages exceeding 1.35 V in $CH_3NH_3Pb(I_{0.8}Br_{0.2})_3$ device. Further optimization of mobility in binary fullerenes electron transport layer can boost the power conversion efficiency as high as 18.6%. We note in particular that the $V_{oc}$-fill factor product is > 1.085 V, which is the highest value reported for halide perovskites with this band gap.

# 1. Introduction

One of the main reasons, why halide perovskite solar cells have attracted so much interest over the past years, is their ability to generate high open-circuit voltages relative to their respective band gaps.[1, 2] Using simple solution-based processing, open-circuit voltages within ~ 60 mV of the radiative limit have been experimentally realized in solar cells with efficiencies exceeding 20%.[3-7] Approaching the radiative limit that closely requires non-radiative recombination processes to be very slow compared to most other semiconductors. This requirement of slow recombination must be valid in the bulk of the material[8] but also at grain boundaries and at interfaces to charge-transport layers. The fact that recombination can be extremely slow even when charge-transport layers are attached to the absorber layer[4, 9-12] is most remarkable and allows combining high open-circuit voltages and thereby high luminescence quantum efficiencies with good fill factors and efficient charge extraction.[13] As shown in **Figure 1**a, high open-circuit voltages relative to the band gap have been shown in halide perovskite solar cells so far mainly for the range of band gaps of 1.5 eV to 1.65 eV. For wider band gaps, approaching the radiative limit has so far been less successful due to increased recombination losses.[14, 15] This is unfortunate, because in particular the band gaps around 1.7 eV to 1.8 eV are highly relevant for making efficient tandem cells based on low band gap absorbers such as Si,[16-19] $Cu(In,Ga)Se_2$[20, 21] or Sn-based low band gap perovskites.[22, 23] One reason for the increased recombination losses for wider-band gap absorbers is halide segregation that leads to stability problems for perovskites containing higher concentrations of Br than about 20%.[15, 24-27] Another obvious but less discussed challenge is the choice of electron and hole transport materials that have to be adapted to the energy levels of the absorber material.[12, 15, 28]

Achieving high open-circuit voltages in perovskite solar cells not only requires well passivated surfaces but also the right choice of charge transfer layer (CTL) for a good energy level alignment.[12] In inverted (anode-illuminated) perovskite solar cells, fullerenes are nearly always used as electron transport layers and high efficiency solar cells with fullerene electron transport based on $C_{60}$ and [6,6]-phenyl-$C_{61}$-butyric acid methyl ester (PCBM) layers have been demonstrated for a range of band gaps.[4, 29-33] However, fullerenes have also restrictions in so far that their energy levels cannot easily be tuned over a wide range. The electron affinity of fullerenes changes with the number and type of adducts attached to the $C_{60}$ cage with multiadducts leading to lower electron affinities relative to monoadducts such as PCBM or even $C_{60}$.[34-36] For instance, for the classical MAPbI$_3$ composition, PCBM provided a good band alignment, while $C_{60}$ (higher electron affinity) leads to additional recombination losses.[37] For higher band gap perovskites with lower perovskite electron affinities, the lowest-unoccupied molecular-orbital (LUMO) of PCBM becomes too low (electron affinity too high) to achieve a perfect band alignment.[38] However, alternatives such as the indene-$C_{60}$ bisadduct (ICBA) with higher-lying LUMO [28, 38, 39] suffers from increased energetic disorder due to the high number of isomers (i.e. there are many different ways of attaching two indene groups to one $C_{60}$ cage). This increased disorder leads to a lower mobility[39, 40] and has previously led to a generally worse electronic properties in organic polymer:fullerene solar cells.[41-44] Therefore, it is currently unclear how to design the electron transport material for perovskite absorbers with a wider range of band gaps.

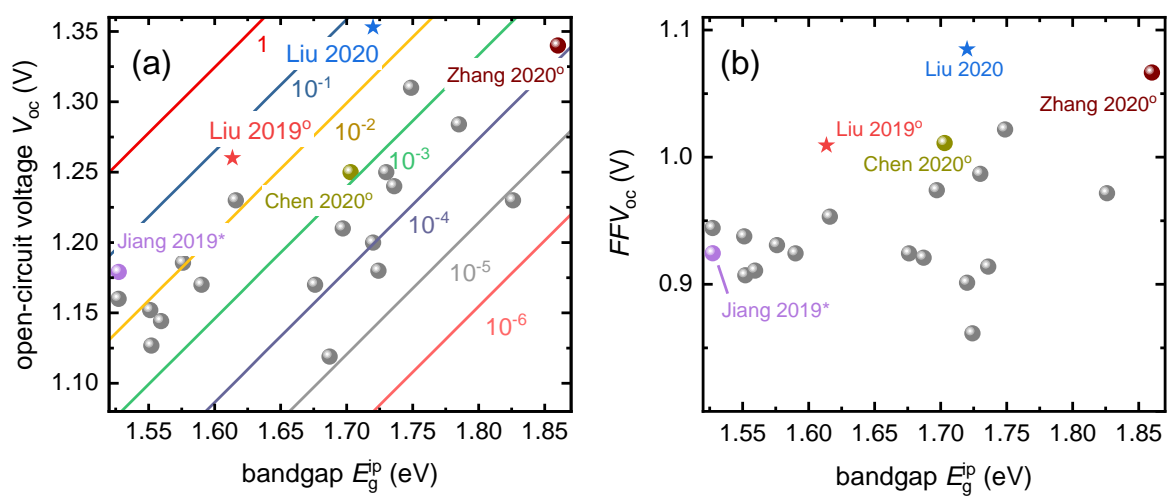

**Figure 1.** (a) Open circuit voltages $V_{oc}$ vs. band gap. The gray spheres giving an overview of the broad spectrum of different band gap perovskites. The band gaps were calculated by the inflection point of the external quantum efficiency. The lines show the external luminescence quantum efficiency ($10^{-6}$-1) calculated with a step function. The red star represents the result of Liu 2019[4] the blue star is the device with the highest $V_{oc}$ of our series with electron transport layer of CMC+ICBA. (b) Product of the fill factor and open circuit voltage $FF \times V_{oc}$ vs. band gap. We used the down sweep for our best cell with $FF = 81.2\%$ and $V_{oc}$=1.35V leading to a product of $FF \times V_{oc}$=1.085 V.

**Figure 1**a shows the $V_{oc}$ vs the band gap of a variety of perovskite solar cells. The colored lines represent the external luminescence quantum efficiency $Q_{e,lum}$ that the solar cells need to have to fulfil the relation[45]

$$q(V_{oc}^{rad} - V_{oc}) = q\Delta V_{oc} = -kT\ln(Q_{e,lum}) \qquad (1)$$

with $V_{oc}^{rad}$ the radiative open-circuit voltage, $q$ the elementary charge, $k$ the Boltzmann constant and $T$ the cell temperature. The band gaps were all calculated by the inflection point of the external (photovoltaic) quantum efficiency to enable better comparison between the data points.[46] The most notable achievements among wide-band gap perovskites are cells based on compositions such as $Cs_{0.17}FA_{0.83}Pb(I_{0.6}Br_{0.4})_3$[15], $Cs_{0.05}(FA_xMA_y)_{0.95}Pb(I_{0.76}Br_{0.24})_3$[47], $Cs_{0.17}FA_{0.83}Pb(I_{0.6}Br_{0.4})_3$[48] and $CsPb(I_{0.75}Br_{0.25})_3$[49]. The band gaps vary from 1.69 eV[47] to 1.86 eV[49] with efficiencies up to 20.1%, at a band gap of 1.69 eV and a $V_{oc}$= 1.21 V[47]. The $V_{oc}FF$ product, a proxy for the voltage at the maximum power point, plotted as a function of $E_g$ for the same devices is shown in **Figure 1**b. The device by Zhang [49] shows a high $V_{oc}$ and high $V_{oc}FF$ product, but a large voltage loss $\Delta V_{oc}$. By applying eq. 1 we can estimate the external luminescence quantum efficiency of the solar cell in ref. [49] to be $Q_{e,lum}$ ~ 0.01%. It is important to stress that due to additional recombination at interfaces to charge transfer layers, the calculated $Q_{e,lum}$ of the devices is much smaller than the external luminescence quantum efficiency of well passivated perovskite films, exceeding $Q_{e,lum}$=35%.[50]

In contrast to Zhang[49], our champion device (blue star) shows low voltage losses leading to high luminescence quantum efficiency $Q_{e,lum}$~ 5% which is comparable with low band gap devices like Jiang 2019[3] (violet) and Liu 2019[4] (red star) (see **Figure 1**a). Furthermore, the champion device shows a high $V_{oc}FF$ product exceeding 1.08 V (see **Figure 1**b). The corresponding solar cell is based on a methylammonium lead iodide-bromide absorber prepared using lead-acetate ($Pb(CH_3COO)_2$), methylammonium iodide (MAI) and methylammonium bromide (MABr) as precursors. This absorber layer is combined with poly(triaryl amine) (PTAA) as a hole transport layer (HTL) and different combinations of fullerenes with varied LUMO as ETLs. We show that open circuit voltages of 1.35 V and efficiencies up to 18.6% can be achieved by tuning the energy level and optimizing the mobility of the fullerenes. We test three different fullerenes, namely PCBM, CMC ($C_{60}$-fused N-Methylpyrrolidine-

m-C$_{12}$-phenyl)[51, 52] and ICBA and all binary combinations of these as electron transport layers (ETLs). ICBA is known to have a roughly 200 meV[53] lower electron affinity as compared to PCBM and has therefore previously been used in higher band gap perovskites.[28] CMC is also a monoadduct fullerene but with a longer side chain and has been reported to have a slightly higher electron affinity (40 meV) than PCBM.[52] While we find high open-circuit voltages > 1.29 V for all of these ETLs, substantial differences in all device parameters are still observed. We find that the best efficiencies and open-circuit voltages up to 1.35 V are possible using a combination of CMC and ICBA to form a binary layer.

## 2. Results
### 2.1. Device Performance

For optimizing wide-band gap CH$_3$NH$_3$Pb(I$_{0.8}$Br$_{0.2}$)$_3$ layers grown on PTAA for highest possible open-circuit voltage, lead acetate based perovskite precursors were chosen because they have been shown[4] to yield smooth, low defect-density perovskite layers which enable the highest reported $V_{oc}$ for perovskite layers with a band gap around 1.6 eV. In order to increase the band gap of the perovskite layer, we doped 20% MABr into CH$_3$NH$_3$PbI$_3$ (MAPI) layer to produce perovskite absorber layers with a wide-band gap of 1.72 eV. For bulk passivation, a previous study[54] has shown that addition of MAI into the perovskite precursor solution can lead to passivation of grain boundaries and substantially increased photoluminescence lifetimes . Our study confirms this by achieving continuous improvement in open-circuit voltages when the excess concentration of MAI is increased from 1.67 mol% to 3.33 mol% in CH$_3$NH$_3$Pb(I$_{0.8}$Br$_{0.2}$)$_3$ precursor solution (**Figure S4**).

**Figure 2**a shows a schematic of our device composed of indium tin oxide (ITO), a 16 nm-thin PTAA as the hole conductor, CH$_3$NH$_3$Pb(I$_{0.8}$Br$_{0.2}$)$_3$ (~280 nm) as the absorber, fullerenes shown in **Figure 2**b (~25 nm) as the electron transport material, as well as bathocuproine (BCP) (~8 nm) and Ag (80 nm) finishing up the cathode. **Figure 2**c shows the current- voltage curve (JV) of the champion device, with the blend of CMC and ICBA (1:1, by weight, sic passim) as electron transport layer. The measurements were recorded at a scan speed of 100 mV/s with a class AAA solar simulator under forward (dashed line) and reverse (solid line) scan conditions. As reported before[4], in order to reach the high open-circuit voltages, all devices have to be photoactivated either by measuring several current

voltage curves or by being kept under open-circuit conditions for 10~40 min depending on the choice of fullerenes by the white light LED (**Figures S2** and **S3**). There could be many reasons for the improvement by light soaking. Possible reasons could be strain reduction in the crystal[55], electronic doping of the fullerene ETLs[56] or movement of the halide ions, leading to defect passivation in the perovskite.[57, 58] In this paper we will not further analyze the cause of the activation effect, but focus on the electrical and optical characterization of the activated cells and layer stacks. During activation, the $J_{sc}$ slightly decreases causing a 1mA/cm$^2$ difference between the $J_{sc}$ computed by integrating the EQE (**Figure S9**) and the $J_{sc}$ measured with the solar simulator. It is important to mention, that all our cells show an increase in $V_{oc}$ if measured again two days after the first measurement (**Figure S5**). Our best cell shows a $V_{oc}$=1.35 V measured on day four after production.

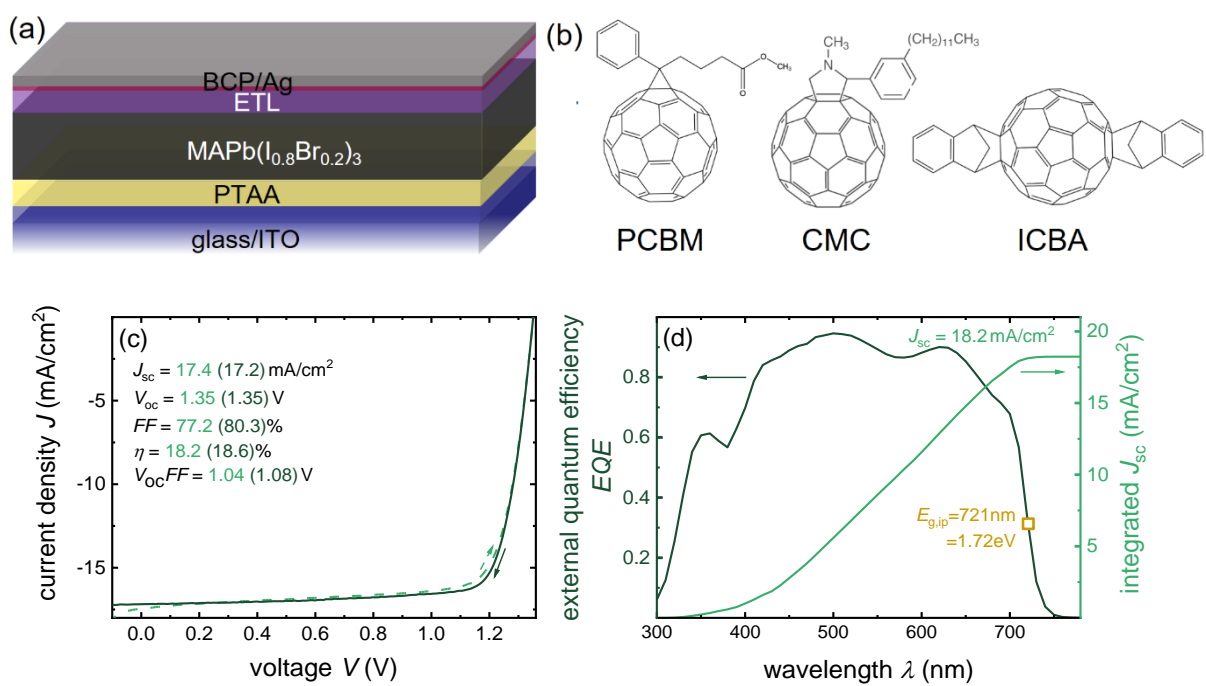

**Figure 2:** (a) Schematic of the device stack. The electron transport layer (ETL) varies between different fullerene derivates and combination of those (b) Structure of PCBM, CMC and ICBA (c) Illuminated current voltage curves of the best device measured four days after production. (d) External quantum efficiency EQE corrected for reflection of the cover glass of the sample holder. The integrated $J_{sc}$=18.2 mA/cm$^2$ is by 1 mA/cm$^2$ higher than the measured $J_{sc}$ due to current loss during activation **(Figure S9).**

## 2.2. Electron Transport Layer Variations

In order to achieve the high $V_{oc}$, we varied the type of fullerenes used as ETL. In the following we want to show the impact of different fullerenes and their blends on the characteristic values of the cells. **Figure 3**a and **3**c show the illuminated *JV*-curves of the best devices, all reaching $V_{oc}$>1.29 V. The statistical data in Figure 3b gives a brief overview of the impact of different ETLs on the cell performance, divided in (from left to right, top to bottom) $J_{sc}$, $V_{oc}$, *FF* and $\eta$ for the pure fullerenes and blends. The box contains 50% of the data points, the antenna shows minimum and maximum value, the bar gives the median and the point is the mean value. The $J_{sc}$ is highest for the cells with PCBM and decreases by mixing with CMC or ICBM. While the cells with pure CMC and ICBA have lower $J_{sc}$, the CMC:ICBA blend shows a similar high $J_{sc}$ as pure PCBM. The $V_{oc}$ shows the opposite trend, PCBM and its blends reach $V_{oc}$>1.33 V while CMC, ICBA and their blend increase the $V_{oc}$ up to 1.35 V. With regard to $J_{sc}$ and $V_{oc}$ CMC and ICBA behave similar, but looking at the *FF* we notice a reduced performance for ICBA and PCBM:ICBA blend. Surprisingly, the blend of CMC and ICBA results in an increase of *FF* compared to both pure materials, the difference between the mean fill factor for CMC, $\Delta FF = 3.4\%$ and for ICBA is even $\Delta FF = 12.4\%$. Thus, the $V_{oc}FF$ product is highest for CMC:ICBA and – combined with the good $J_{sc}$ – this ETL leads to the highest efficiency $\eta = 18.9\%$. The other devices reach efficiencies of approximately 17% with the exception of the ICBA-based device, where the poor *FF* leads to a low efficiency of $\eta < 16\%$. Each triangle in **Figure 3**d illustrates the trend between the pure fullerenes (corner) and their blends in $V_{oc}$, *FF* and $V_{oc}FF$ product, respectively. The values are shown for the cells with the highest efficiency and color-coded from blue for low $V_{oc}$, *FF* and $V_{oc}FF$ product to red for the maximum of $V_{oc}$, *FF* and $V_{oc}FF$.

Thus, to summarize the findings from **Figure 3**, we find that the relatively high electron affinity of PCBM indeed leads to the lowest open-circuit voltages in the series, verifying the need for looking into alternative electron transport layer materials. The substitution of PCBM by ICBA enables higher open-circuit voltages thereby verifying the assumption that the energy level alignment at the perovskite-PCBM interface leaves room for improvement. The cells based on ICBA, however, also show a drastically reduced *FF* consistent with the observed charge transport problems of fullerene multiadducts encountered in the past in the context of polymer:fullerene solar cells.[42, 43] The apparent solution for this dilemma is found to be the use of the fullerene monoadduct CMC blended with either PCBM or – with even higher efficiencies – ICBA. Thus, the use of binary blends of different fullerenes allows combining their advantages rather than being limited by the shortcomings of the respective molecules. This is an observation that has also previously been made in the context of organic solar cells, where ternary blends using e.g. two different acceptor molecules combined with one donor have efficiencies that exceed those of the binary blends.[59-61]

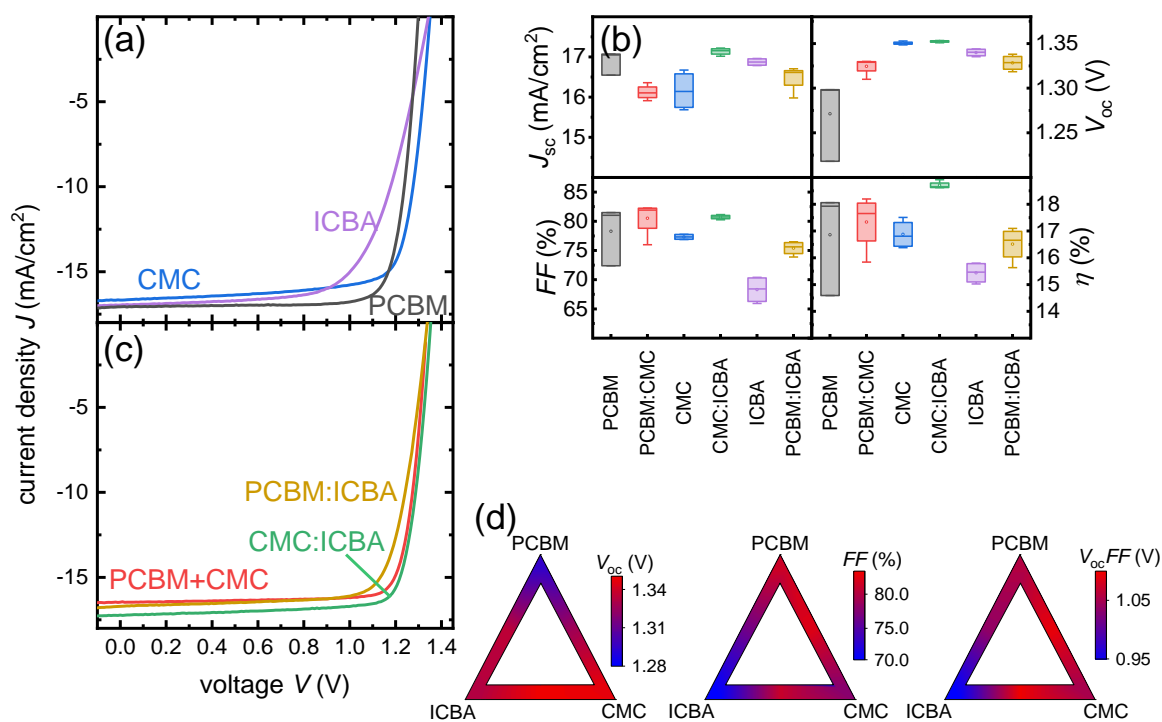

**Figure 3:** (a), (c) Illuminated current voltage curves of cells with different ETLs. Measured on a AAA sun simulator after activation with LED. (b) $J_{sc}$, $V_{oc}$, $FF$ and $\eta$ statistic of the four cells on one sample. The box contains 50% of the data, the antenna shows minimum and maximum values, the bars gives the median (in case of 4 cells the mean value of both median values) and the point gives the mean value. (d) The triangles show the $V_{oc}$, $FF$ and $V_{oc}FF$ product indicated by a color map of the cells with the highest efficiency for each fullerene and the 1:1 blend.

## 2.3. Loss Analysis

In the following, we will briefly study the key loss mechanisms in our solar cells by comparing the performance indicators $J_{sc}$, $V_{oc}$, $FF$ and efficiency to the values of a cell with the same band gap in the Shockley-Queisser model. We use the methodology that we introduced in refs. [62] and [63].

In **Figure 4**a the power density of our champion device is shown as a function of voltage (blue) and is compared with the power density of a cell with the same band gap (1.72 eV) with an ideal $FF$ (red), ideal $V_{oc}$ and ideal $FF$ (green) and finally the power density-voltage curve of the cell in the Shockley-Queisser model (yellow). **Figure 4**a emphasizes that we optimized the cells for high $V_{oc}$ and high $FF$. Thus, the difference between ideal $J_{sc}$ and measured $J_{sc}$ is the highest but can be explained by the low absorber thickness that gives the best electronic quality and highest open-circuit voltages. The $V_{oc}$ is close to the Shockley-Queisser limit (green curve) for a band gap of 1.72 eV. As described in ref. [63] a better way to compare the quality of the $V_{oc}$ is by comparing the measured and the radiative open-circuit voltage $V_{oc}^{rad}$. We calculated $V_{oc}^{rad}$ by fitting an Urbach tail to the external quantum efficiency (**Figure 4**b). The difference $\Delta V_{oc} = V_{oc}^{rad} - V_{oc} = 70$ mV confirms the good $V_{oc}$ comparable to Liu 2019 [4] ($\Delta V_{oc} = 64$ mV) and Jiang 2019[3] ($\Delta V_{oc} = 67$ mV). Notably, this loss is compatible with an external luminescence quantum efficiency $Q_{e,lum} = 6.56\%$ which is the highest reported value in the band gap region around 1.7 eV.

**Figure 4**c shows the ratio of different losses for our champion device with a 1:1 blend of CMC and ICBA and 5 other fullerenes or fullerene blends as ETL. The percentage of SQ efficiency is plotted on a logarithmic axis, i.e. the relative size of the boxes is independent of their position in the total loss bar. The normalized efficiency can be written as

$$\frac{\eta_{real}}{\eta_{SQ}} = F_{sc} \frac{V_{oc}^{real}}{V_{oc}^{SQ}} \frac{FF_0(V_{oc}^{real})}{FF_0(V_{oc}^{SQ})} F_{FF}^{res} , \qquad (2)$$

where $F_{sc} = J_{sc}/J_{sc}^{SQ}$ is the photocurrent loss (yellow **Figure 4**c). Note that the $FF$ depends on the $V_{oc}$. Hence the loss in $FF$ is divided in the loss in maximum fill factor that is due to the loss in $V_{oc}$, i.e. $FF_0(V_{oc}^{real})/FF_0(V_{oc}^{SQ})$ (lilac **Figure 4**c), and the loss in fill factor that is due to the series resistance and the ideality factor, which we denote as $F_{FF}^{res} = FF_{real}/FF_0(V_{oc}^{real})$ (red **Figure 4**c). Regardless of the fullerene, the loss in $J_{sc}$ is the highest loss. All cells show a small loss in calculated *FF* due to a small difference in $V_{oc}^{SQ}$ and measured $V_{oc}$. The resistive loss in *FF* varies for the different ETL. In case of ICBA $F_{FF}^{res}$ is particularly high suggesting a reduced conductivity of the ICBA layer. The loss in $V_{oc}$ can be divided in the difference between the actual absorption coefficient and the assumed step function in SQ (blue **Figure 4**c) and non-radiative loss (green **Figure 4**c). The perovskite and thus the band gap does not change and hence the ratio $V_{oc}^{rad}/V_{oc}^{SQ}$ is nearly the same for all cells. The ratio $V_{oc}^{real}/V_{oc}^{rad}$ is highest for cells which contain PCBM in the ETL and lowest for cells with either ICBA, CMC or ICBA:CMC as ETL. Our champion device based on the CMC:ICBA blend ETL shows a moderate normalized efficiency $\eta_{real}/\eta_{SQ} = 0.66$, but our non-radiative $V_{oc}$ loss is comparable with high efficiency devices based on much lower band gaps, e.g. Jiang 2019 ref [3]. Figure 4d illustrates the exceptionally low voltage losses of our solar cells compared to the voltage losses in other wide band gap perovskite devices that typically exceed those of lower band gap perovskites.[15, 49, 64, 65]

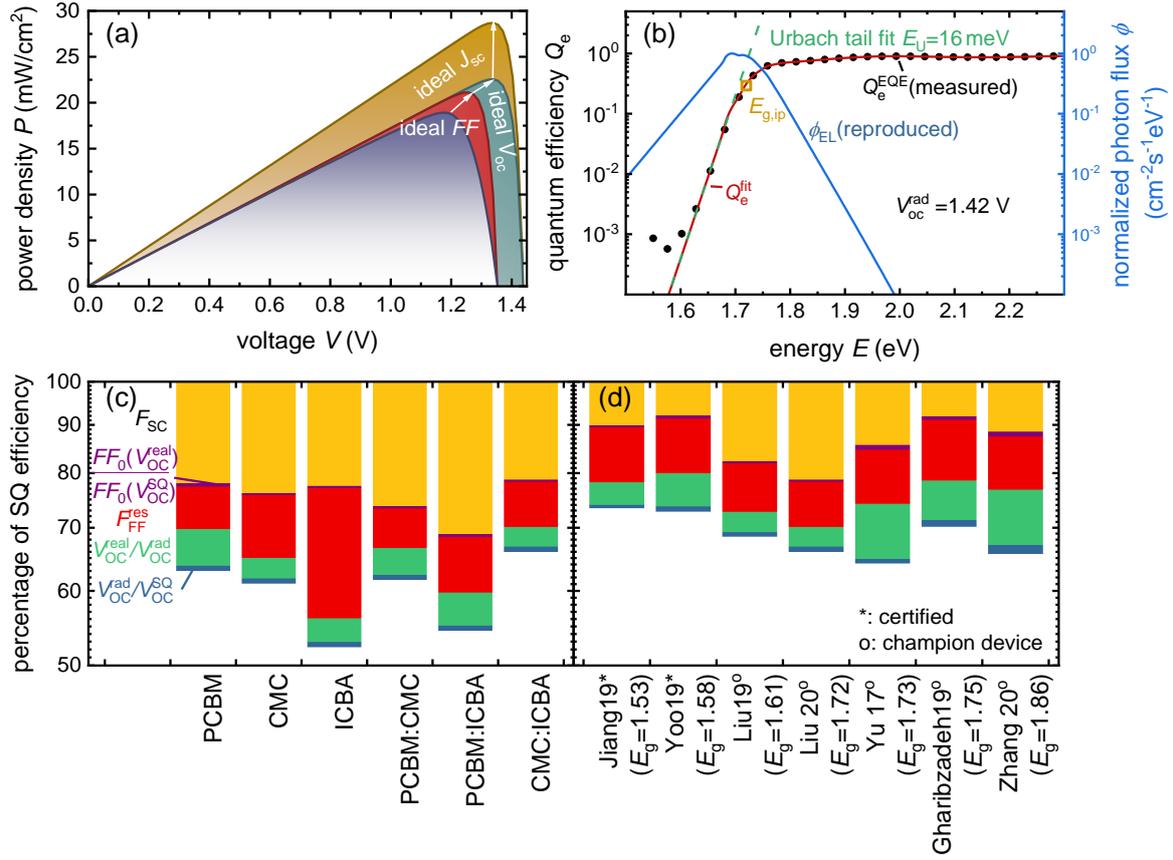

**Figure 4:** Analysis of losses. (a) Power density $P$ of the best cell (blue) compared to the ideal cell in different steps of ideality, ideal fill factor (red), Shockley-Queisser $V_{oc}$ (green) and complete SQ model (yellow). The biggest improvement could be achieved by increasing $J_{sc}$. (b) The radiative limit $V_{oc}^{rad}$=1.42 V is calculated from the product of photovoltaic quantum efficiency and black body spectrum as described e.g. in ref.[45] . The photovoltaic quantum efficiency is extended by fitting an Urbach tail of $E_U$=16 meV to the measured quantum efficiency $Q_e^{EQE}$ as explained in ref.[63] Because the $V_{oc}^{rad}$ should be independent of the ETL, we show the data from the cell with the CMC:ICBA blend as ETL. (c) The percentage of Shockley-Queisser efficiency for cells with different ETLs separated in current loss ($F_{sc}$), fill factor losses ($F_{FF}^{res}$, $FF_0(V_{oc}^{real})/FF_0(V_{oc}^{SQ})$) and radiative and non-radiative voltage loss ($V_{oc}^{rad}/V_{oc}^{SQ}$, $V_{oc}^{real}/V_{oc}^{rad}$). (d) Percentage of Shockley-Queisser efficiency for perovskite cells with different band gaps.

## 2.4. Interfacial Recombination

The most obvious finding from our previous analysis of device data is the observation of high open-circuit voltages and low non-radiative recombination losses. This implies that our process yields a high-quality bulk material which only shows minimal losses due to interface recombination. Interface recombination is affected both by the energy-level alignment[66-68] at the interface between absorber and both charge transport layers and by its kinetics that are typically expressed in terms of surface or interface recombination velocities.[69] Here, we will first study the kinetics of recombination using transient photoluminescence measurements and subsequently the energy-level alignment of the different fullerenes by a combination of ultraviolet photoelectron spectroscopy (UPS) and photothermal deflection spectroscopy (PDS).

### 2.4.1. Recombination kinetics

We measured transient photoluminescence with time-correlated single-photon counting (TCSPC) on half cells without the BCP and silver contact. This was done to reduce the impact of capacitive effects that would be expected from electrode charging and discharging in complete devices.[70] In order to analyze the data, we calculate the differential decay time by first fitting the Tr-PL decays and then differentiating the fits using

$$\tau_{\text{diff,HLI}} = -2 \left( \frac{d ln(\phi)}{dt} \right)^{-1} . \qquad (3)$$

Where $\phi$ is the luminescence the factor of 2 originates from the assumption of high-level injection (HLI), i.e. $\phi \propto n^2$ with charge-carrier concentration $n$. The resulting differential decay times are plotted as a function of the quasi Fermi-level splitting $\Delta E_F$ which is proportional to $\ln(\phi)$ as computed in equation S5. The differential lifetimes as shown in **Figures 5**c, d. show a rapidly increasing lifetime at large quasi Fermi-level splitting, i.e high luminescence (short times), and a small rise at intermediate values of $\Delta E_F$. At smaller values of $\Delta E_F$ (long times) the decay times differ for the different fullerenes. CMC:ICBA (green) and PCBM:ICBA (gold) nearly saturate in a slightly sloped plateau, PCBM (gray) and ICBA (lilac), continue rising but with a smaller slope, PCBM:CMC (red) saturates after a steep rise and CMC (blue) increases fast. It should be mentioned that the steady state PL shows a peak shift (**Figure S13**) for PCBM:CMC, which could be the reason for the difference to the other mixed fullerenes. The decays are generally comparably slow as those measured on similar MAPI samples as presented e.g. in refs. [4] and [69].

The first regime (short times) can be affected by charge extraction to the charge transport layers and the ITO electrode as well as by radiative and Auger recombination, leading to relatively low decay times.[71, 72] The intermediate and late regime should be dominated by either bulk or interface recombination with an approximately constant lifetime, i.e. some type of defect assisted SRH process. In contrast to the other fullerenes and blends the differential decay time of CMC increases even further, at lower values of $\Delta E_F$ (long times), representing the flat parts of the Tr-PL decays shown in **Figure 5**a. These regimes are representative of the part of the original data with the lowest signal to noise ratio, where the Tr-PL data approaches its background (noise) level. For the other ETLs we cut off the regime with lowest signal to noise as described in **Figure S16**. In addition, the long time regime can be affected by slow processes such as capacitive discharges of electrodes or contact layers that reinject charge into the perovskite absorber. Therefore, we conclude that the region most representative of the information of interest, namely all types of SRH recombination, is the flat or intermediate region. If we analyze the data of the 1:1 blend of CMC and ICBA at the point where the derivative of the differential decay time has a minimum, we obtain differential decay times of $\tau_{eff} \approx 0.75$ µs at $\Delta E_F = 1.46$ eV. From these decay times, we can estimate surface recombination velocities

$$S(\Delta E_F) = 2d \left( \frac{1}{\tau_{\text{diff,HLI}}(\Delta E_F)} - \frac{1}{\tau_{\text{bulk}}} \right) \tag{4}$$

of $S = 17$ cm/s up to $67$ cm/s by variation of the bulk recombination time from $\tau_{\text{bulk}} = 1$ µs to $10$ µs and a thickness $d = 280$ nm (**Figure 5**e, f). These values of the surface recombination velocity are slightly higher than what was previously estimated for the PTAA/MAPI interface[4] based on Tr-PL measurements but slightly lower than previous estimates[69, 71] for recombination at the MAPI/PCBM interface.

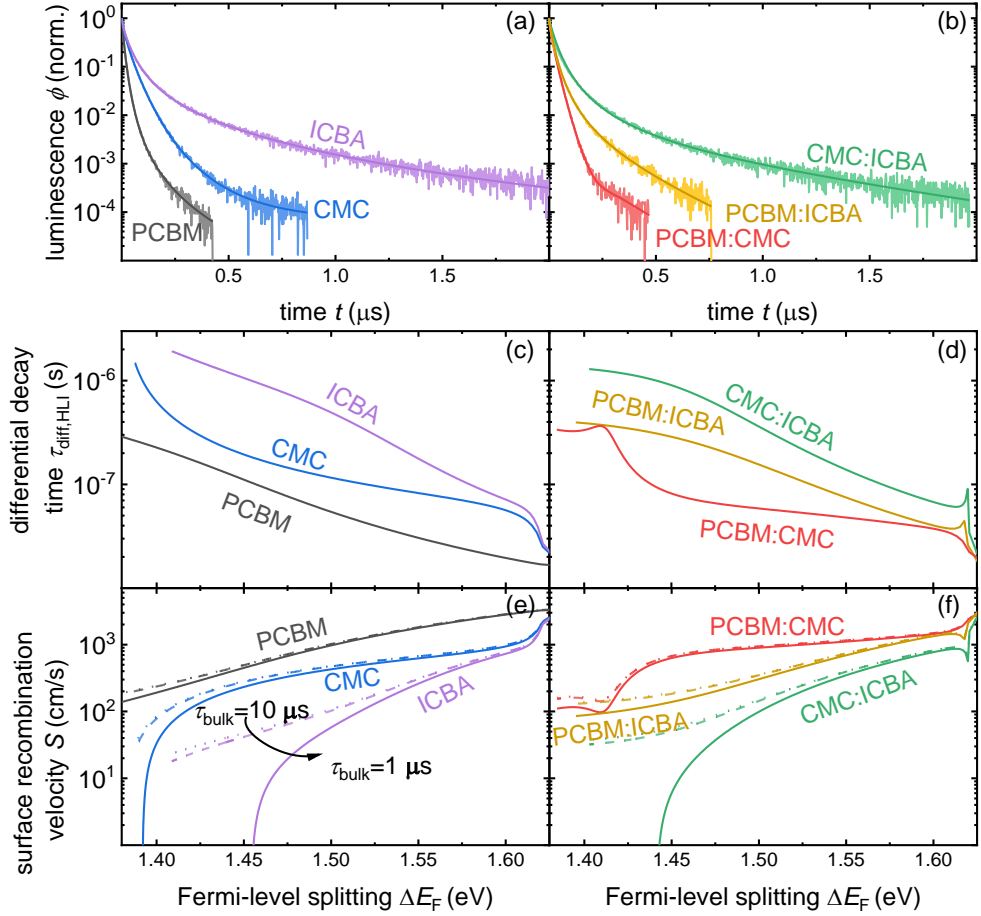

**Figure 5:** (a) Normalized transient photoluminescence $\phi$ of glass/ITO/PTAA/Perovskite/ETL samples with pure and (b) mixed fullerenes measured with time correlated single-photon counting (TCSPC). In order to analyze the data, we fit two exponential functions to the normalized data (darker lines). (c), (d) We computed the differential decay time $\tau_{\text{diff,HLI}}$ by using eq. (3) and plotted the result vs. Fermi-level splitting $\Delta E_F$. Samples using an ETL made from a (1:1) blend of CMC and ICBA show the highest lifetime of $\tau > 1$ µs. (e), (f) Surface recombination velocity $S$ computed with eq (3) for bulk lifetime $\tau_{\text{bulk}} = 10$ µs to 1 µs.

### 2.4.2. Energy Level Alignment and $V_{oc}$-Losses

In addition to surface quality the energy alignment between perovskite and CTL limits the $V_{oc}$ [12, 69] Large offsets e.g. in the conduction band at the absorber-ETL interface would lead to an interfacial band gap at this interface that is substantially lower than the bulk band gap. If recombination is efficient via the interface (i.e. electrons in the ETL can efficiently recombine with holes in the perovskite), the interface may strongly deteriorate the $V_{oc}$. Therefore, characterization of energy levels is crucial for a complete understanding of interfacial recombination in a solar cell. Typically, UPS is used to measure the work function $\Phi$ and the valence-band edge $E_v$ at the surface of the samples. Given that we are interested in the LUMO positions of our ETLs, we have to combine UPS data with data from optical spectroscopy to obtain the band gap and thereby the electron affinity $\chi$.

We derived the work function and the valence-band edge with UPS measurements as shown in **Figure S19** for all fullerenes and blends on glass/ITO/PTAA/perovskite/ETL stacks. We activated the half cells before the UPS measurement by illumination under a white LED lamp as described in SI. **Figure 6**c illustrates the difference between the ionization energy $E_i = E_{vac} - E_V$ of PCBM and the other fullerenes. We determined a smaller $E_i$ for ICBA of 160 meV, which is in good agreement with the offset (200 meV) measured by ref. [53]. The relevant quantity for band alignment with the perovskite is however the electron affinity as it determines electron extraction and recombination of electrons in the ETL with holes in the perovskite. However, fullerenes do not show a clear absorption onset that could be used to determine the band gap but feature a weakly absorbing absorption feature at around 1.75eV and then a slow rise until strong absorption is visible in the blue and UV parts of the spectrum. This has led to the situation electron affinities determined from combination of UPS and absorption spectroscopy vary wildly in the literature[40, 52, 73-76]. In order to circumvent this challenge, we again only compare relative changes in band gap that we determine from PDS measurements analyzed at three different constant absorption coefficients $\alpha = 1 \times 10^4 \text{ cm}^{-1}$, $\alpha = 5 \times 10^3 \text{ cm}^{-1}$ and $\alpha = 1 \times 10^3 \text{ cm}^{-1}$. The arithmetic mean of the three optical band gap differences is shown in **Figure 6**b. We observe that the optical band gap of CMC is approximately 40 meV bigger than the bandgap of PCBM while the other fullerenes show slightly lower band gaps than PCBM.

**Figure 6**d shows the relative position of the LUMO of the different ETLs compared to PCBM determined from the values in 6b and c and the median $V_{oc}$ from the solar cell characterization. The error bars of $V_{oc}$ represent the datapoints within 1.5xIQR, were the interquartile range (IQR) represents the region of 50% of the data points. (**Figure S10**). PCBM and PCBM:CMC share the same electron affinity, while the other blends and fullerenes have a lower electron affinity. The LUMO of ICBA is closest to vacuum, fitting to the high $V_{oc}$. However, the correlation between $V_{oc}$ and $\Delta\chi$ is not distinct enough to solely explain the high $V_{oc}$. Instead, we propose that the combination of the optimization of the energy level alignment and slight improvements in the interface passivation suggested by the Tr-PL measurements qualitatively explain the trends in $V_{oc}$. A quantitative prediction of the trends of $V_{oc}$ with ETL may be a target for future studies but has so far not been attempted due to the difficulties in the numerical simulations of Tr-PL decays and the experimental uncertainties in measuring energy level alignment with accuracies of a tens of mV.

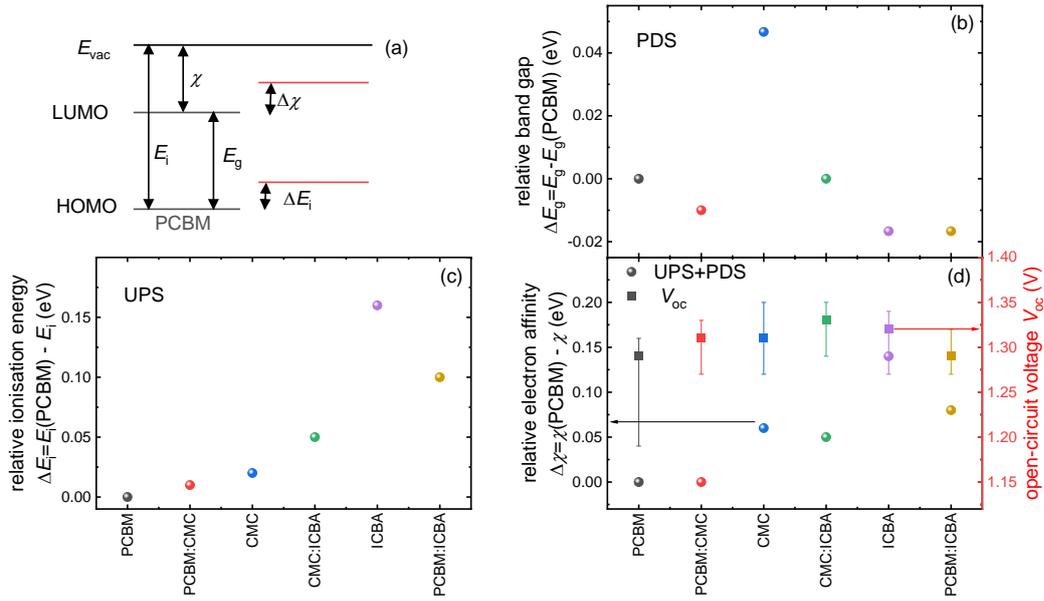

**Figure 6:** (a) Schematic band diagram of PCBM compared to the other ETLs (b) Band gap difference $\Delta E_g$ between PCBM and the other ETL materials, measured with PDS and total transmittance and reflectance spectroscopy. (c) Ionization energy $E_i = E_{vac} - E_V$ relative to the ionization energy of PCBM. The work function $\Phi$ and the valence band energy relative to the Fermi-level energy $E_F = 0$ eV were derived from UPS measurements. (d) Difference between electron affinity of PCBM and the other ETLs. The electron affinity was computed with the PDS measurements and UPS measurements and the median $V_{oc}$ with the error bar representing the datapoints within 1.5×IQR (**Figure S12).**

## 2.5. Fill-factor losses

As observed in Figure 4c and d, the fill factor losses are still substantial and even at fill factors ~ 80% generally exceed the losses in $V_{oc}$ in the cells discussed here. The same is true for most high efficiency perovskite solar cells reported in literature and stands out relative to more mature technologies such as Si or GaAs where these losses are substantially lower.[63] These losses partly originate from the high ideality factors observed even in high efficiency perovskite solar cells[13, 77] and from the resistive losses induced by the relative low conductivity and low permittivity of the organic charge transport layers.[78] In the following, we will therefore study the general distribution of *FF* losses in more detail and analyze in particular resistive losses due to low mobilities in the ETLs via space charge limited current measurements.

### 2.5.1. Analysis of current-voltage curves

In order to distinguish between resistive contributions to the fill factor loss and those based on ideality factor $n_{id}$ higher than one, we use a simple method to determine the series resistance $R_s$ by comparing illuminated and dark *JV*-curve in the 4$^{th}$ quadrant.[79, 80] When neglecting the parallel resistance and assuming the superposition principle[81, 82] holds, the diode equation

$$J_l = J_0 \times \left( \exp\left( \frac{q(V_l - J_l R_s)}{n_{id} kT} \right) - 1 \right) - J_{sc} \qquad (6)$$

with *k* the Boltzmann constant and *T* the cell temperature and $J_0 = J_{sc} \times \exp(-qV_{oc}/kT)$ describes the *JV*-curve of a solar cells. In the dark the diode equation is reduced to

$$J_d = J_0 \times \left( \exp\left( \frac{q(V_d - J_d R_s)}{n_{id} kT} \right) - 1 \right), \qquad (7)$$

where the *JV*-curve is independent of light induced effects. The $J_{sc}/V_{oc}$-curve, i.e. $J_{sc}$ and $V_{oc}$ measured at different illumination intensities is independent of $R_s$,

$$J_{sc} = J_0 \times \left( \exp\left( \frac{qV_{oc}}{n_{id} kT} \right) - 1 \right) \qquad (8)$$

since the current is zero at $V_{oc}$. **Figure 7**a and b show the illuminated *JV*-curve (blue) the dark *JV*-curve shifted by $J_{sc}$ (yellow), the $J_{sc}/V_{oc}$-curve shifted by $J_{sc}$ (red) and the ideal *JV*-curve (green), i.e. $R_s = 0\ \Omega\text{cm}^2$ and $n_{id} = 1$ for a PCBM cell. The difference between illuminated *JV*-curve and dark *JV*-curve divided by $J_{sc}$ leads to the series resistance

$$R_S = (V_d - V_l)/(J_d - J_l) = (V_d - V_l)/J_{sc}, \qquad (9)$$

the mean series resistance for the PCBM cell is $R_s \approx 4.9\ \Omega\text{cm}^2$ (**Figure 7**c). In **Figure 7**d and e we plotted the power density vs. voltage for the different *JV*-curves in a similar manner to **Figure 4**a. The maximum of the power density curve gives us the efficiency of our solar cell. The difference between illuminated and $J_{sc}/V_{oc}$-curve reveals the fill factor loss $\Delta FF_{R_s} = FF_l - FF_{J_{sc}V_{oc}} = 3.7\%$ due to the series resistance in the cell.. The difference between $J_{sc}/V_{oc}$-curve and ideal power density curve gives us the loss due to non-ideal diode behavior, i.e. an ideality factor $n_{id} > 1$.[83] The ideality factor for PCBM is in the range of roughly $n_{id} \sim 1.7$ to 1.9 (**Figure S25**). The fill factor loss due to non-ideal diode behavior is slightly larger than the loss due to $R_s$ $\Delta FF_{n_{id}} = FF_{id} - FF_{J_{sc}V_{oc}} = 5.9\%$ (see **Table 1**).

**Table 1:** $J_{sc}$, $V_{oc}$, $FF$ and $\eta$ for a solar cell based on an PCBM ETL compared to $FF$ and $\eta$ from an ideal $JV$-curve with the same $J_{sc}$ and $V_{oc}$ but $R_s = 0$ and $n_{id} = 1$ and a curve with zero series resistance (but non-ideal $n_{id}$), derived from the measurement of the $J_{sc}/V_{oc}$ curves shown in Figure 8a and b.

|  | $J_{sc}$ (mA/cm$^2$) | $V_{oc}$ (V) | $FF_{JscVoc}$ (%) | $FF_{id}$ (%) | $\eta$ (%) | $\eta_{JscVoc}$ (%) | $\eta_{id}$ (%) | $\Delta FF_{R_s}$ (%) | $\Delta FF_{n_{id}}$ (%) |
|---|---|---|---|---|---|---|---|---|---|
| PCBM | 16.4 | 1.30 | 84.8 | 90.7 | 17.3 | 18.1 | 19.4 | 3.7 | 5.9 |

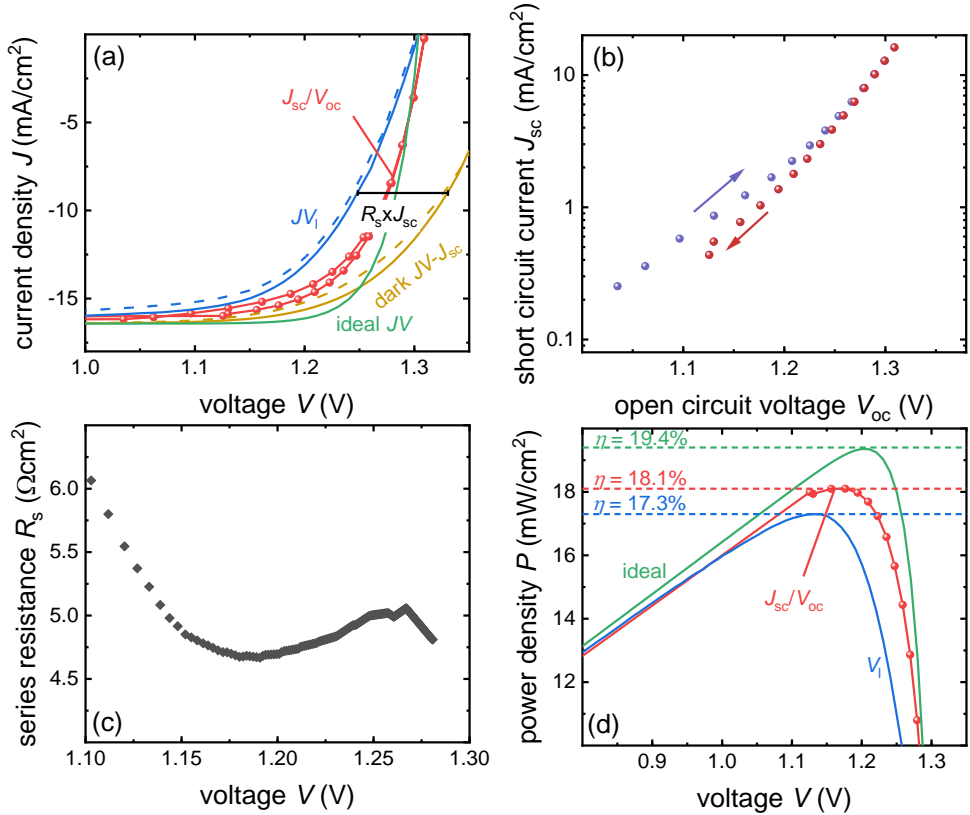

**Figure 7:** (a) Illuminated current voltage curve (blue) forward (dashed line) and reversed (solid line) scan direction, of a PCBM cell measured with an AAA sun simulator. Short circuit current density vs. open circuit voltage measured on a white LED subtracted by the short circuit density at 1 Sun (red). Dark $JV$-curve subtracted by $J_{sc}$ (green) and ideal $JV$-curve following eq. 5. (b) short-circuit current plotted against open-circuit voltage of the PCBM cell at different light intensities. From low to intensity ~ 0.03 sun to roughly 1 sun (open symbols) and back to ~ 0.03 sun (filled symbols) (c) The difference between illuminated $JV$ and dark $JV$ leads to the series resistance $R_s$ for PCBM (d) Power density of the illuminated, dark, $J_{sc}/V_{oc}$ and ideal curve. The maximum power density is equal to the efficiency of the solar cells.

### 2.5.2. Space-charge limited current measurements

The observation of improved *FF* for the binary ETLs compared to pure fullerene ETLs as shown in **Figure 3**d, suggests that the choice of fullerene or fullerene blend affects the conductivity of the ETL which in turn could modulate the resistive losses caused by the ETL. Hence, we studied the difference in electron mobility $\mu_e$ across the different ETLs using space-charge-limited current (SCLC) measurements. Therefore, we constructed electron only devices, similar to the cell stack. As shown in **Figure 8**a, we used ZnO nanoparticles and the standard BCP/Ag contact also used in the solar cells as electron injecting and extracting contact for the electron-only devices. In SCLC measurements, the current should follow the Mott-Gurney law[84]

$$J = \frac{9}{8}\epsilon_0\epsilon_r\mu_e\frac{V^2}{d^3}, \qquad (10)$$

where $\epsilon_0$ is the vacuum permittivity, $\epsilon_r = 3$ the assumed relative permittivity and $d$ the thickness of the fullerenes and fullerene blends if certain conditions are met. These conditions include most notably the absence of diffusion – an assumption that is typically violated if there are either charged defects or asymmetric injection barriers present in the device. In order to distinguish between both effects, we measured the dark $JV$-curve from -4 V to 4 V with the difference between forward and reverse bias being indeed substantial (see **Figure S21**).[85, 86] To minimize the influence of diffusion on the data, we use the curve obtained by injecting electrons at the BCP/Ag electrode and extracting them at the ZnO electrode and limit ourselves to voltages $V > 1$ V.

**Figure 8**a shows a double-logarithmic representation of this part of the data shown as $J \times d^3$ vs. voltage showing a transition from a region affected by diffusion to a region that is approximately following the Mott-Gurney law for higher voltages. Note that assuming constant permittivity among the different fullerenes, higher values of $J \times d^3$ should always correspond to higher mobilities according to eq. (10) the Mott-Gurney law can only be applied in the region of the flat slope.[87] All samples except the ICBA-based sample show a transition from a high slope at lower voltages (diffusion limited region) to a smaller slope at higher voltages that approaches the value of 2 expected from the Mott-Gurney law. The curves for the different fullerenes are approximately parallel suggesting that they differ essentially in mobility. On the contrary the ICBA-based sample shows a substantially higher slope than 2 throughout the whole voltage region suggesting that energetic disorder affects charge transport.[88] **Figure 8**a leads to the conclusion that PCBM:CMC and CMC have the highest mobility while ICBA shows the lowest mobility of all six fullerenes and – given the higher slope of current vs. voltage —is affected by energetic disorder.[88]

To extract the mobility, we simulated the complete *JV*-curve with the drift-diffusion simulation solver *Advanced Semiconductor Analysis* (ASA) to compute the mobility as described in SI. **Figure 8**b shows the result of numerically fitting the curves and extracting the electron mobility $\mu_e$ for the different ETLs. This electron mobility is then plotted as a function of fill factor *FF*. The ZnO nanoparticles form a rough layer which leads to uncertainties in the thickness of the fullerene layers that are reflected in the error bar of the mobility. The fill factor is the median of all measured cells at a AAA sun simulator on day 4 after activation, while the error bars represent 1.5×IQR. With the exception of the CMC sample that shows a fairly good mobility relative to a modest *FF*, we observe a monotonous increase of *FF* with electron mobility from ICBA (worst *FF*) to PCBM:CMC (best *FF*). 1D-drift-diffusion simulations of *JV*-curves (**Figure S23**) confirm the electron mobility trend shown in **Figure 8**b for all ETLs but CMC. Note that the CMC sample is inhomogeneous (**Figure S24**), thus the thickness measurements may have a larger error bar, or the material properties could be different to those properties present in a cell with much thinner CMC layer.

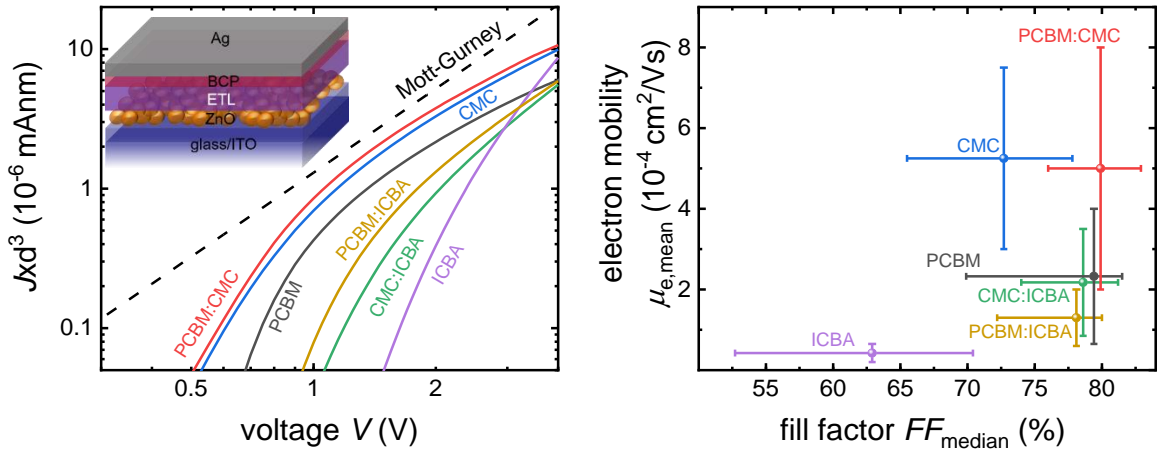

**Figure 8:** (a) Space-charge-limited current (SCLC) $J \times d^3$ vs. voltage for the bias direction where electrons are injected from the BCP/Ag and extracted at the ZnO interface (Comparison between forward and reverse direction can be found in supplementary figure S21). For high voltages, the curve should follow the Mott-Gurney law (Equation 4) Inset: Schematic of the electron-only device stack using BCP/Ag und ZnO as the electron injecting materials. (b) Electron mobility from ASA simulation (see SI) plotted against the median *FF* of all cells measured with AAA sun simulator on day four. The error bar for the mobility represents the uncertainty in thickness due to surface roughness. The error bar of *FF* represents the datapoints within 1.5×IQR (**Figure S12**).

## 3. Conclusion

Reaching high open-circuit voltage is one of the main challenges for wide band gap perovskite solar cells. Using $CH_3NH_3Pb(I_{0.8}Br_{0.2})_3$ with a band gap of $E_g = 1.72$ eV and a blend of CMC:ICBA as electron transport layer we managed to fabricate solar cells with a high $V_{oc} = 1.35$ eV leading to a non-radiative loss of only $\Delta V_{oc} = 70$ mV without compromising in $FF$. Using transient photoluminescence spectroscopy and a combination of ultraviolet photoelectron spectroscopy and photothermal deflection spectroscopy, we were able to assign the high $V_{oc}$ to low interfacial recombination and partly improved energy alignment. In addition, we studied the losses in fill factor, which are substantial given that corresponding $FF$ in the SQ model is nearly 91% for a $V_{oc}$ of 1.35 V. By using a comparison of measured current-voltage curves and suns-$V_{oc}$ measurements, we find that the $FF$ losses are distributed approximately equally between losses due to the ideality factor $n_{id} > 1$ and due to resistive losses. The resistive losses most likely originate from the finite conductivity of the charge transport layers. Space-charge-limited current measurements show that the electron mobility in the ETL materials varies for all fullerenes and fullerene blends, showing a monotonic correlation between $FF$ and electron mobility that holds for all samples but one. This suggests that further pathways towards higher efficiencies can be based on higher conductivities and mobilities in the electron transport layers combined with a better understanding of ideality-factor losses. The latter losses may be reduced by either further approaching the radiative limit for example by better surface quality or by moving the most recombination active parts of the device into low level injection.

### Acknowledgments

The authors acknowledge support from the Helmholtz Association for funding via the PEROSEED project. Z.L and J.S want to thank Oliver Thimm for the PDS measurements. The authors also thank the Initiative and Networking Fund of the Helmholtz Association for funding of the JOSEPH cluster system via the Helmholtz Energy Materials Characterization Platform (HEMCP).

**Literature**

[1]   W. Tress, *Advanced Energy Materials* **2017**, *7*, 1602358.
[2]   J. P. Correa-Baena, M. Saliba, T. Buonassisi, M. Grätzel, A. Abate, W. Tress, A. Hagfeldt, *Science* **2017**, *358*, 739.
[3]   Q. Jiang, Y. Zhao, X. Zhang, X. Yang, Y. Chen, Z. Chu, Q. Ye, X. Li, Z. Yin, J. You, *Nature Photonics* **2019**, *13*, 460.
[4]   Z. Liu, L. Krückemeier, B. Krogmeier, B. Klingebiel, J. A. Márquez, S. Levcenko, S. Öz, S. Mathur, U. Rau, T. Unold, T. Kirchartz, *ACS Energy Letters* **2019**, *4*, 110.
[5]   M. Jeong, I. W. Choi, E. M. Go, Y. Cho, M. Kim, B. Lee, S. Jeong, Y. Jo, H. W. Choi, J. Lee, J.-H. Bae, S. K. Kwak, D. S. Kim, C. Yang, *Science* **2020**, *369*, 1615.
[6]   H. Lu, Y. Liu, P. Ahlawat, A. Mishra, W. R. Tress, F. T. Eickemeyer, Y. Yang, F. Fu, Z. Wang, C. E. Avalos, B. I. Carlsen, A. Agarwalla, X. Zhang, X. Li, Y. Zhan, S. M. Zakeeruddin, L. Emsley, U. Rothlisberger, L. Zheng, A. Hagfeldt, M. Grätzel, *Science* **2020**, *370*, eabb8985.
[7]   G. Kim, H. Min, K. S. Lee, D. Y. Lee, S. M. Yoon, S. I. Seok, *Science* **2020**, *370*, 108.
[8]   I. L. Braly, D. W. deQuilettes, L. M. Pazos-Outon, S. Burke, M. E. Ziffer, D. S. Ginger, H. W. Hillhouse, *Nature Photonics* **2018**, *12*, 355.
[9]   M. Abdi-Jalebi, Z. Andaji-Garmaroudi, S. Cacovich, C. Stavrakas, B. Philippe, J. M. Richter, M. Alsari, E. P. Booker, E. M. Hutter, A. J. Pearson, S. Lilliu, T. J. Savenije, H. Rensmo, G. Divitini, C. Ducati, R. H. Friend, S. D. Stranks, *Nature* **2018**, *555*, 497.
[10]   A. Al-Ashouri, A. Magomedov, M. Roß, M. Jošt, M. Talaikis, G. Chistiakova, T. Bertram, J. A. Márquez, E. Köhnen, E. Kasparavičius, S. Levcenco, L. Gil-Escrig, C. J. Hages, R. Schlatmann, B. Rech, T. Malinauskas, T. Unold, C. A. Kaufmann, L. Korte, G. Niaura, V. Getautis, S. Albrecht, *Energy & Environmental Science* **2019**, *12*, 3356.
[11]   M. Stolterfoht, C. M. Wolff, J. A. Márquez, S. Zhang, C. J. Hages, D. Rothhardt, S. Albrecht, P. L. Burn, P. Meredith, T. Unold, D. Neher, *Nature Energy* **2018**, *3*, 847.
[12]   M. Stolterfoht, P. Caprioglio, C. M. Wolff, J. A. Márquez, J. Nordmann, S. Zhang, D. Rothhardt, U. Hörmann, Y. Amir, A. Redinger, L. Kegelmann, F. Zu, S. Albrecht, N. Koch, T. Kirchartz, M. Saliba, T. Unold, D. Neher, *Energy & Environmental Science* **2019**, *12*, 2778.
[13]   M. Stolterfoht, M. Grischek, P. Caprioglio, C. M. Wolff, E. Gutierrez‐Partida, F. Peña‐Camargo, D. Rothhardt, S. Zhang, M. Raoufi, J. Wolansky, *Advanced Materials* **2020**, 2000080.
[14]   C. Chen, Z. Song, C. Xiao, D. Zhao, N. Shrestha, C. Li, G. Yang, F. Yao, X. Zheng, R. J. Ellingson, C.-S. Jiang, M. Al-Jassim, K. Zhu, G. Fang, Y. Yan, *Nano Energy* **2019**, *61*, 141.
[15]   Y. Yu, C. Wang, C. R. Grice, N. Shrestha, D. Zhao, W. Liao, L. Guan, R. A. Awni, W. Meng, A. J. Cimaroli, K. Zhu, R. J. Ellingson, Y. Yan, *ACS Energy Letters* **2017**, *2*, 1177.
[16]   Z. Yu, M. Leilaeioun, Z. Holman, *Nature Energy* **2016**, *1*, 16137.
[17]   T. Leijtens, K. A. Bush, R. Prasanna, M. D. McGehee, *Nature Energy* **2018**, *3*, 828.
[18]   J. Xu, C. C. Boyd, Z. J. Yu, A. F. Palmstrom, D. J. Witter, B. W. Larson, R. M. France, J. Werner, S. P. Harvey, E. J. Wolf, W. Weigand, S. Manzoor, M. F. A. M. van Hest, J. J. Berry, J. M. Luther, Z. C. Holman, M. D. McGehee, *Science* **2020**, *367*, 1097.
[19]   D. Kim, H. J. Jung, I. J. Park, B. W. Larson, S. P. Dunfield, C. Xiao, J. Kim, J. Tong, P. Boonmongkolras, S. G. Ji, F. Zhang, S. R. Pae, M. Kim, S. B. Kang, V. Dravid, J. J. Berry, J. Y. Kim, K. Zhu, D. H. Kim, B. Shin, *Science* **2020**, *368*, 155.
[20]   Q. Han, Y.-T. Hsieh, L. Meng, J.-L. Wu, P. Sun, E.-P. Yao, S.-Y. Chang, S.-H. Bae, T. Kato, V. Bermudez, Y. Yang, *Science* **2018**, *361*, 904.

[21]   S. Gharibzadeh, I. M. Hossain, P. Fassl, B. A. Nejand, T. Abzieher, M. Schultes, E. Ahlswede, P. Jackson, M. Powalla, S. Schäfer, *Advanced Functional Materials* **2020**, *30*, 1909919.
[22]   G. E. Eperon, T. Leijtens, K. A. Bush, R. Prasanna, T. Green, J. T.-W. Wang, D. P. McMeekin, G. Volonakis, R. L. Milot, R. May, A. Palmstrom, D. J. Slotcavage, R. A. Belisle, J. B. Patel, E. S. Parrott, R. J. Sutton, W. Ma, F. Moghadam, B. Conings, A. Babayigit, H.-G. Boyen, S. Bent, F. Giustino, L. M. Herz, M. B. Johnston, M. D. McGehee, H. J. Snaith, *Science* **2016**, *354*, 861.
[23]   J. Tong, Z. Song, D. H. Kim, X. Chen, C. Chen, A. F. Palmstrom, P. F. Ndione, M. O. Reese, S. P. Dunfield, O. G. Reid, J. Liu, F. Zhang, S. P. Harvey, Z. Li, S. T. Christensen, G. Teeter, D. Zhao, M. M. Al-Jassim, M. F. A. M. van Hest, M. C. Beard, S. E. Shaheen, J. J. Berry, Y. Yan, K. Zhu, *Science* **2019**, *364*, 475.
[24]   S. Mahesh, J. M. Ball, R. D. J. Oliver, D. P. McMeekin, P. K. Nayak, M. B. Johnston, H. J. Snaith, *Energy & Environmental Science* **2020**, *13*, 258.
[25]   E. T. Hoke, D. J. Slotcavage, E. R. Dohner, A. R. Bowring, H. I. Karunadasa, M. D. McGehee, *Chemical Science* **2015**, *6*, 613.
[26]   D. J. Slotcavage, H. I. Karunadasa, M. D. McGehee, *ACS Energy Letters* **2016**, *1*, 1199.
[27]   A. Sadhanala, F. Deschler, T. H. Thomas, S. E. Dutton, K. C. Goedel, F. C. Hanusch, M. L. Lai, U. Steiner, T. Bein, P. Docampo, *the journal of physical chemistry letters* **2014**, *5*, 2501.
[28]   S. Wheeler, D. Bryant, J. Troughton, T. Kirchartz, T. Watson, J. Nelson, J. R. Durrant, *The Journal of Physical Chemistry C* **2017**, *121*, 13496.
[29]   Y. Lin, B. Chen, F. Zhao, X. Zheng, Y. Deng, Y. Shao, Y. Fang, Y. Bai, C. Wang, J. Huang, *Advanced Materials* **2017**, *29*, 1700607.
[30]   Y. Shao, Y. Yuan, J. Huang, *Nature Energy* **2016**, *1*, 15001.
[31]   Y. Shao, Z. Xiao, C. Bi, Y. Yuan, J. Huang, *Nature Communications* **2014**, *5*, 5784.
[32]   D. Luo, W. Yang, Z. Wang, A. Sadhanala, Q. Hu, R. Su, R. Shivanna, G. F. Trindade, J. F. Watts, Z. Xu, *Science* **2018**, *360*, 1442.
[33]   C. Zuo, L. Ding, *Advanced Energy Materials* **2017**, *7*, 1601193.
[34]   Y. He, H.-Y. Chen, J. Hou, Y. Li, *Journal of the American Chemical Society* **2010**, *132*, 1377.
[35]   M. Lenes, G.-J. A. H. Wetzelaer, F. B. Kooistra, S. C. Veenstra, J. C. Hummelen, P. W. M. Blom, *Advanced Materials* **2008**, *20*, 2116.
[36]   H. Wang, Y. He, Y. Li, H. Su, *The Journal of Physical Chemistry A* **2012**, *116*, 255.
[37]   C. M. Wolff, F. Zu, A. Paulke, L. P. Toro, N. Koch, D. Neher, *Advanced Materials* **2017**, *29*, 1700159.
[38]   Y. Lin, B. Chen, F. Zhao, X. Zheng, Y. Deng, Y. Shao, Y. Fang, Y. Bai, C. Wang, J. Huang, *Adv Mater* **2017**, *29*.
[39]   P. W. Liang, C. C. Chueh, S. T. Williams, A. K. Y. Jen, *Advanced Energy Materials* **2015**, *5*, 1402321.
[40]   J. Chen, X. Lian, Y. Zhang, W. Yang, J. Li, M. Qin, X. Lu, G. Wu, H. Chen, *Journal of Materials Chemistry A* **2018**, *6*, 18010.
[41]   J. M. Frost, M. A. Faist, J. Nelson, *Advanced Materials* **2010**, *22*, 4881.
[42]   F. Steiner, S. Foster, A. Losquin, J. Labram, T. D. Anthopoulos, J. M. Frost, J. Nelson, *Materials Horizons* **2015**, *2*, 113.
[43]   M. A. Faist, S. Shoaee, S. Tuladhar, G. F. A. Dibb, S. Foster, W. Gong, T. Kirchartz, D. D. C. Bradley, J. R. Durrant, J. Nelson, *Advanced Energy Materials* **2013**, *3*, 744.
[44]   M. A. Faist, P. E. Keivanidis, S. Foster, P. H. Wöbkenberg, T. D. Anthopoulos, D. D. C. Bradley, J. R. Durrant, J. Nelson, *Journal of Polymer Science Part B: Polymer Physics* **2011**, *49*, 45.
[45]   U. Rau, *Physical Review B* **2007**, *76*, 085303.
[46]   U. Rau, B. Blank, T. C. Müller, T. Kirchartz, *Physical Review Applied* **2017**, *7*, 044016.
[47]   F. Peña-Camargo, P. Caprioglio, F. Zu, E. Gutierrez-Partida, C. M. Wolff, K. Brinkmann, S. Albrecht, T. Riedl, N. Koch, D. Neher, *ACS Energy Letters* **2020**.
[48]   S. Gharibzadeh, B. Abdollahi Nejand, M. Jakoby, T. Abzieher, D. Hauschild, S. Moghadamzadeh, J. A. Schwenzer, P. Brenner, R. Schmager, A. A. Haghighirad, L. Weinhardt, U. Lemmer, B. S. Richards, I. A. Howard, U. W. Paetzold, *Advanced Energy Materials* **2019**, *9*, 1803699.
[49]   J. Zhang, Z. Wang, A. Mishra, M. Yu, M. Shasti, W. Tress, D. J. Kubicki, C. E. Avalos, H. Lu, Y. Liu, *Joule* **2020**, *4*, 222.

[50]     D. W. deQuilettes, S. Koch, S. Burke, R. K. Paranji, A. J. Shropshire, M. E. Ziffer, D. S. Ginger, *ACS Energy Letters* **2016,** *1*, 438.
[51]     M. Chikamatsu, K. Kikuchi, T. Kodama, H. Nishikawa, I. Ikemoto, N. Yoshimoto, T. Hanada, Y. Yoshida, N. Tanigaki, K. Yase, *AIP Conference Proceedings* **2001,** *590*, 455.
[52]     D. B. Khadka, Y. Shirai, M. Yanagida, T. Noda, K. Miyano, *ACS Applied Materials & Interfaces* **2018,** *10*, 22074.
[53]     M. A. Faist, T. Kirchartz, W. Gong, R. S. Ashraf, I. McCulloch, J. C. de Mello, N. J. Ekins-Daukes, D. D. C. Bradley, J. Nelson, *Journal of the American Chemical Society* **2012,** *134*, 685.
[54]     D.-Y. Son, J.-W. Lee, Y. J. Choi, I.-H. Jang, S. Lee, P. J. Yoo, H. Shin, N. Ahn, M. Choi, D. Kim, N.-G. Park, *Nature Energy* **2016,** *1*, 16081.
[55]     H. Tsai, R. Asadpour, J.-C. Blancon, C. C. Stoumpos, O. Durand, J. W. Strzalka, B. Chen, R. Verduzco, P. M. Ajayan, S. Tretiak, *Science* **2018,** *360*, 67.
[56]     M. De Bastiani, G. Dell'Erba, M. Gandini, V. D'Innocenzo, S. Neutzner, A. R. S. Kandada, G. Grancini, M. Binda, M. Prato, J. M. Ball, *Advanced Energy Materials* **2016,** *6*, 1501453.
[57]     D. W. DeQuilettes, W. Zhang, V. M. Burlakov, D. J. Graham, T. Leijtens, A. Osherov, V. Bulović, H. J. Snaith, D. S. Ginger, S. D. Stranks, *Nature communications* **2016,** *7*, 1.
[58]     E. Mosconi, D. Meggiolaro, H. J. Snaith, S. D. Stranks, F. De Angelis, *Energy & Environmental Science* **2016,** *9*, 3180.
[59]     P. P. Khlyabich, B. Burkhart, B. C. Thompson, *Journal of the American Chemical Society* **2011,** *133*, 14534.
[60]     R. A. Street, D. Davies, P. P. Khlyabich, B. Burkhart, B. C. Thompson, *Journal of the American Chemical Society* **2013,** *135*, 986.
[61]     D. Baran, R. S. Ashraf, D. A. Hanifi, M. Abdelsamie, N. Gasparini, J. A. Röhr, S. Holliday, A. Wadsworth, S. Lockett, M. Neophytou, C. J. M. Emmott, J. Nelson, C. J. Brabec, A. Amassian, A. Salleo, T. Kirchartz, J. R. Durrant, I. McCulloch, *Nature Materials* **2017,** *16*, 363.
[62]     J.-F. Guillemoles, T. Kirchartz, D. Cahen, U. Rau, *Nature Photonics* **2019,** *13*, 501.
[63]     L. Krückemeier, U. Rau, M. Stolterfoht, T. Kirchartz, *Advanced energy materials* **2020,** *10*, 1902573.
[64]     J. J. Yoo, S. Wieghold, M. C. Sponseller, M. R. Chua, S. N. Bertram, N. T. P. Hartono, J. S. Tresback, E. C. Hansen, J.-P. Correa-Baena, V. Bulović, *Energy & Environmental Science* **2019,** *12*, 2192.
[65]     S. Gharibzadeh, B. Abdollahi Nejand, M. Jakoby, T. Abzieher, D. Hauschild, S. Moghadamzadeh, J. A. Schwenzer, P. Brenner, R. Schmager, A. A. Haghighirad, *Advanced Energy Materials* **2019,** *9*, 1803699.
[66]     S. Wang, T. Sakurai, W. Wen, Y. Qi, *Advanced Materials Interfaces* **2018,** *5*, 1800260.
[67]     P. Schulz, D. Cahen, A. Kahn, *Chemical Reviews* **2019,** *119*, 3349.
[68]     C. M. Wolff, P. Caprioglio, M. Stolterfoht, D. Neher, *Advanced Materials* **2019,** *31*, 1902762.
[69]     J. Haddad, B. Krogmeier, B. Klingebiel, L. Krückemeier, S. Melhem, Z. Liu, J. Hüpkes, S. Mathur, T. Kirchartz, *Advanced Materials Interfaces* **2020,** *7*, 2000366.
[70]     D. Kiermasch, A. Baumann, M. Fischer, V. Dyakonov, K. Tvingstedt, *Energy & Environmental Science* **2018,** *11*, 629.
[71]     B. Krogmeier, F. Staub, D. Grabowski, U. Rau, T. Kirchartz, *Sustainable Energy & Fuels* **2018,** *2*, 1027.
[72]     T. Kirchartz, J. A. Márquez, M. Stolterfoht, T. Unold, *Advanced Energy Materials* **2020,** *10*, 1904134.
[73]     K. Akaike, K. Kanai, H. Yoshida, J. y. Tsutsumi, T. Nishi, N. Sato, Y. Ouchi, K. Seki, *Journal of Applied Physics* **2008,** *104*, 023710.
[74]     J.-Y. Jeng, K.-C. Chen, T.-Y. Chiang, P.-Y. Lin, T.-D. Tsai, Y.-C. Chang, T.-F. Guo, P. Chen, T.-C. Wen, Y.-J. Hsu, *Advanced Materials* **2014,** *26*, 4107.
[75]     Z.-L. Guan, J. B. Kim, Y.-L. Loo, A. Kahn, *Journal of Applied Physics* **2011,** *110*, 043719.
[76]     R. Nakanishi, A. Nogimura, R. Eguchi, K. Kanai, *Organic Electronics* **2014,** *15*, 2912.
[77]     P. Caprioglio, C. M. Wolff, O. J. Sandberg, A. Armin, B. Rech, S. Albrecht, D. Neher, M. Stolterfoht, *Advanced Energy Materials* **2020,** *10*, 2000502.

[78] O. J. Sandberg, J. Kurpiers, M. Stolterfoht, D. Neher, P. Meredith, S. Shoaee, A. Armin, *Advanced Materials Interfaces* **2020,** *7*, 2000041.
[79] T. C. M. Müller, B. E. Pieters, U. Rau, T. Kirchartz, *Journal of Applied Physics* **2013,** *113*, 134503.
[80] M. Wolf, H. Rauschenbach, *Advanced Energy Conversion* **1963,** *3*, 455.
[81] F. A. Lindholm, J. G. Fossum, E. L. Burgess, *IEEE Transactions on Electron Devices* **1979,** *26*, 165.
[82] S. J. Robinson, A. G. Aberle, M. A. Green, *Journal of Applied Physics* **1994,** *76*, 7920.
[83] P. Rostan, U. Rau, V. Nguyen, T. Kirchartz, M. Schubert, J. Werner, *Solar energy materials and solar cells* **2006,** *90*, 1345.
[84] N. F. Mott, R. W. Gurney, **1948**.
[85] T. Kirchartz, *Beilstein journal of nanotechnology* **2013,** *4*, 180.
[86] J. A. Röhr, *Physical Review Applied* **2019,** *11*, 054079.
[87] J. A. Röhr, T. Kirchartz, J. Nelson, *Journal of Physics: Condensed Matter* **2017,** *29*, 205901.
[88] P. Mark, W. Helfrich, *Journal of Applied Physics* **1962,** *33*, 205.